\documentclass[a4paper,11pt]{article}
\pdfoutput=1 

\usepackage{jheppub} 

\usepackage[T1]{fontenc} 

\pdfoutput=1
\usepackage{amsmath,amssymb}
\usepackage{amsfonts}
\usepackage{bm,bbm}
\usepackage{graphicx}
\usepackage{mathrsfs}
\usepackage{slashed}
\usepackage{booktabs}
\usepackage{tabu}
\usepackage[dvipsnames]{xcolor}
\usepackage[normalem]{ulem}
\usepackage{tikz}
\usepackage{soul}

\usepackage{hyperref}
\hypersetup{colorlinks,citecolor= blue,linkcolor= blue, urlcolor=blue}

\usepackage{xcolor}
\usepackage{ulem}
\usepackage{array}
\usepackage{verbatim}
\usepackage{epsfig}
\usepackage{multirow}

\newcommand{\be}{\begin{equation}}
\newcommand{\ee}{\end{equation}}
\newcommand{\ba}{\begin{array}}
\newcommand{\ea}{\end{array}}
\newcommand{\bea}{\begin{eqnarray}}
\newcommand{\eea}{\end{eqnarray}}

\usepackage{ulem,fancyvrb}
\usepackage{xcolor}

\title{New channel to search for dark matter at Belle II}

\author[a,b,c]{Jinhan Liang,}
\author[a,d]{Zuowei Liu}
\author[a]{and Lan Yang}

\affiliation[a]{Department of Physics, Nanjing University, Nanjing 210093, China}
\affiliation[b]{Guangdong Provincial Key Laboratory of Nuclear Science, Institute of Quantum Matter, South China Normal University, Guangzhou 510006, China}
\affiliation[c]{Guangdong-Hong Kong Joint Laboratory of Quantum Matter, Southern Nuclear Science Computing Center, South China Normal University, Guangzhou 510006, China}
\affiliation[d]{CAS Center for Excellence in Particle Physics, Beijing 100049, China}

\emailAdd{jinhanliang@smail.nju.edu.cn}
\emailAdd{zuoweiliu@nju.edu.cn}
\emailAdd{lanyang@smail.nju.edu.cn}

\abstract{

We propose a new ``disappearing positron track'' channel at Belle II to search for dark matter, in which a positron that is produced at the primary interaction vertex scatters with the electromagnetic calorimeter to produce dark matter particles. Such scatterings can occur via either annihilation with atomic electrons, or the bremsstrahlung process with target nuclei. The main backgrounds are due to photons and neutrons that are produced in the same scatterings and then escape detection. We require a large missing energy and further veto certain activities in the KLM detector to suppress such backgrounds. To illustrate the sensitivity of the new channel, we consider a new physics model where dark matter interacts with the standard model via a dark photon, which decays predominantly to dark matter; we find that our proposed channel can probe some currently unexplored parameter space, 
surpassing both the mono-photon channel at Belle II 
and the NA64 constraints. }

\begin{document}

\maketitle
\flushbottom

\section{Introduction}

Dark matter (DM) is one of the most fundamental and longstanding questions in modern physics 
\cite{Bertone:2016nfn}. 
However, the particle property of DM remains elusive, despite the overwhelming 
evidence of its gravitational effects 
\cite{Alexander:2016aln, Battaglieri:2017aum,Arbey:2021gdg,Bertone:2016nfn}. 
Particle colliders are one of the most powerful tools to search for DM. 
The leading DM signature studied at colliders is the so-called mono-X channel 
where DM are usually produced at the primary collision vertex 
of the collider accompanied by 
a standard model (SM) particle X 
\cite{Birkedal:2004xn, Feng:2005gj, Beltran:2010ww, Bai:2010hh, Petriello:2008pu}.
Over the years, a plethora of mono-X processes have been studied, 
with $X$ being photon \cite{Fox:2011fx, Gershtein:2008bf, Abdallah:2015uba, Birkedal:2004xn}, 
jet \cite{Feng:2005gj, Beltran:2010ww,Bai:2010hh,Fox:2011pm,Rajaraman:2011wf, Papucci:2014iwa}, 
top \cite{ Andrea:2011ws, Agram:2013wda, Boucheneb:2014wza}, 
bottom \cite{Lin:2013sca, Izaguirre:2014vva}, 
$Z/W$ \cite{Carpenter:2012rg, Bell:2012rg, Bell:2015rdw, Bell:2015sza, Petriello:2008pu, Haisch:2016usn, Bai:2012xg}, 
or Higgs \cite{Petrov:2013nia, Carpenter:2013xra, Berlin:2014cfa, Ghorbani:2016edw, No:2015xqa}.

In this paper,  
we propose a new channel to search for DM at colliders 
where DM are
produced in collisions between SM particles 
and the detector, instead of at the primary collision vertex 
of the collider. 
For concreteness, in this analysis 
we take as an example the Belle II experiment, 
the electron-positron collider operated at {SuperKEKB}. 
Belle II is expected to accumulate at least 50 ab$^{-1}$ data 
and has a hermetic electromagnetic calorimeter (ECL) 
\cite{Belle-II:2018jsg}, 
making it an ideal machine 
to search
for light DM as well as  
other new light particles 
\cite{Hauth:2018fgp, Liang:2019zkb, Araki:2017wyg, Zhang:2020fiu, Kang:2021oes,Duerr:2019dmv, Duerr:2020muu, Ferber:2022ewf, Belle-II:2020jti, Izaguirre:2016dfi, Jho:2019cxq, Belle-II:2019qfb, Liang:2021kgw, Dolan:2017osp, Bandyopadhyay:2022klg}.

\begin{figure}[htbp]
\begin{centering}
\includegraphics[width=0.4 \textwidth]{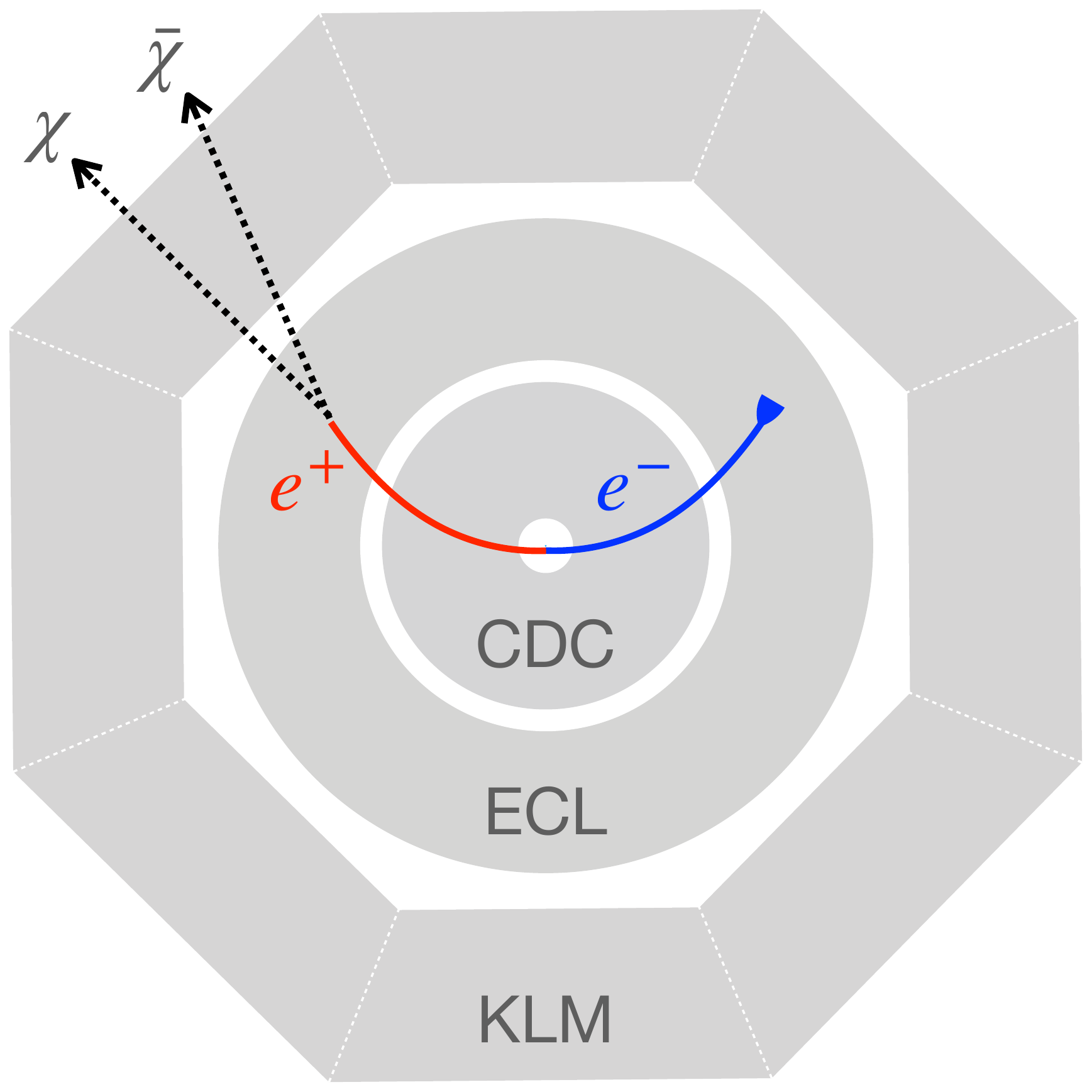}
\caption{Schematic view of the signal event in  
the ``disappearing positron track'' channel, 
in the transverse plane of the Belle II detector.}
\label{fig:detector}
\end{centering}
\end{figure}

\begin{figure}[htbp]
\begin{centering}
\includegraphics[width=0.75 \textwidth]{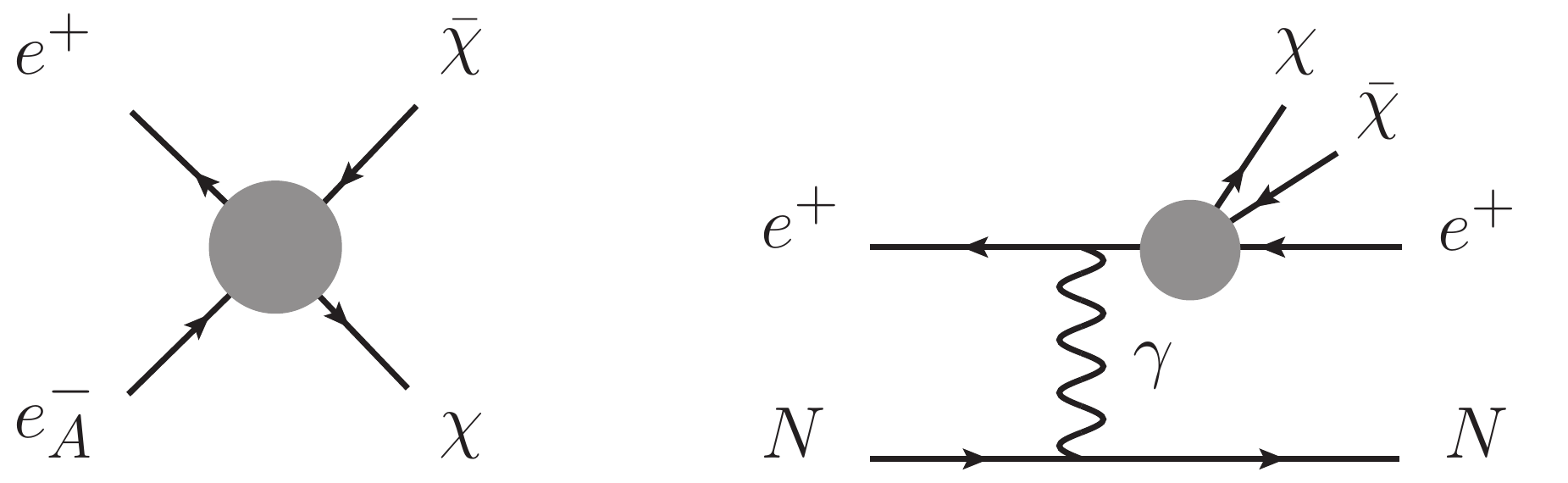}
\caption{Feynman diagrams of the annihilation process 
$e^+ e_A^- \to \bar{\chi}\chi$ (left), 
and the bremsstrahlung process 
$e^+ N \to e^+ N \bar{\chi}\chi$ (right). 
The process with $\bar \chi \chi$ radiated from 
the initial state 
$e^+$ is included in the analysis, but not shown here.} 
\label{fig:Feynmandig}
\end{centering}
\end{figure}

Electrons and positrons are copiously produced 
at Belle II via the Bhabha scattering process, 
leading to ${\cal O}(10^{12})$ positrons expected 
with 50 ab$^{-1}$. 
These final state positrons can then 
interact
with  
the ECL detector to produce DM, 
as shown in Fig.~(\ref{fig:detector}).
The interactions between the positron and the ECL
can occur either via 
annihilations with atomic electrons in the ECL 
(the annihilation process)
\be
e^+ + e_A^- \to \bar \chi + \chi, 
\ee
or via 
scatterings with target nuclei in the ECL  
(the bremsstrahlung process)
\be 
e^+ + N \to  e^+ +N+ \bar{\chi} + \chi, 
\ee
where $\chi$ is the DM particle,   
$e_A^-$ is the atomic electron in the ECL, 
and $N$ is the target nucleus, 
which can be either Cs or I in the ECL.  
The Feynman diagrams of these two 
processes are shown in Fig.\ \eqref{fig:Feynmandig}. 
The DM particles then escape the Belle II detectors, 
resulting in a missing energy signature. 
We note that this channel is analogous to that in 
electron fixed-target experiments 
(e.g., NA64 \cite{Andreev:2021fzd}), 
with the ECL detector as the target.

Unlike DM 
produced at the primary collision vertex, 
the missing energy in this new channel is preceded by a charged track in the 
central drift chamber (CDC) 
and a small amount of energy deposited in the ECL.  
We thus refer to this new channel as the  ``disappearing positron track'' channel.\footnote{We note that disappearing tracks  
have been searched for at the LHC in new physics studies; 
see e.g., Refs.~\cite{ATLAS:2022rme, CMS:2023mny}. 
However, to our knowledge, 
the mechanism that leads to the disappearing track 
in this analysis has not been studied before. 
Moreover, the signature presented in this analysis has a unique event topology, 
which is distinct from various LHC studies.}
Moreover, the disappearing positron is accompanied by an electron that has an opposite momentum 
to the positron track  
in the center of mass (c.m.) frame, 
a clear CDC track, 
and an energy deposition in the ECL that is consistent with 
the Bhabha scattering.

The collisions between positrons and the ECL 
can also produce photons and neutrons, 
which have a non-zero probability to penetrate  
all the sub-detectors of Belle II without leaving a trace, 
mimicking the signal events of DM. 
We find that a large missing energy in the ECL 
and a veto on a cluster or a track in the 
KLM ($K_L$-muon detector) \cite{Belle-II:2018jsg, Dolan:2017osp}
are instrumental in controlling these backgrounds.

To illustrate the capability of this new channel in 
probing DM, we consider the dark photon (DP) model 
in which the DP predominantly decays into DM. 
The main results are summarized in Fig.~(\ref{fig:BelleIIDP}). 
For the DP mass in the vicinity of 66 MeV, 
the new channel can probe 
unexplored parameter space, 
surpassing both the 
mono-photon channel at Belle II 
and the NA64 experiment. 
We note that, 
despite our efforts to reproduce the Belle II conditions to the best of our ability,
the limits presented in this analysis 
should be revised by Belle II physicists 
using a full simulation of the detectors   
and inputs from data control samples. 
In particular, the various efficiency factors associated 
with the ECL and KLM detectors, which play a pivotal role 
in estimating the background, will be updated with a better understanding  
as the experiment progresses.

This rest of the paper is organized as follows.
In section \ref{sec:channel}, 
we discuss the ``disappearing positron track'' signature at Belle II. 
In section \ref{sec:BG},
we discuss various SM backgrounds for the new signature. 
In section \ref{sec:results}, 
we compute the signal events in the new channel 
for the invisible dark photon model, 
and further compare its sensitivity to other experimental searches, 
including the mono-photon channel at Belle II and 
the missing energy channel at NA64. 
In section \ref{sec:conclusion}, 
we summarize our findings. 
In the appendix, we provide the positron track length distribution  for completeness, 
explore the dependence of the signal on the coupling constant,
present our calculations on the NA64 constraints, 
 discuss the backgrounds due to charged particles,
compute the sensitivity on the scalar mediator model, 
and
explore the sensitivity of the radiative Bhabha process.

\section{Disappearing positron track}
\label{sec:channel}

The energy of the final state positron is measured by the ECL detector, 
which has a barrel region and two endcap regions.
In our analysis, we only consider the positrons in the 
barrel region (with a polar angle between $32.2^\circ$ 
and $128.7^\circ$ in the lab frame), 
due to the following three reasons. 
First,
there are less non-instrumented setups, 
such as magnetic wires and pole tips, 
between ECL and KLM in the barrel region as compared to 
the endcap regions \cite{Ferber}. 
This leads to a better KLM veto efficiency in the barrel region, 
which is essential in controlling the SM background.
Second, 
the barrel region has a better hermiticity: 
Gaps between ECL crystals in the barrel region 
are non-projective to the collision point; 
however, 
some gaps in the endcap regions are projective 
so that particles can escape the ECL detector 
without being noticed when traversing them
\cite{Hearty, Belle:2000cnh}. 
Third, the endcap regions have more beam backgrounds \cite{deJong:2017ofr}.

Although in the signal process 
the positron cannot deposit all its energy 
in the ECL due to the production of DM, 
its transverse momentum can be measured in the CDC with 
a good resolution, e.g., 
$\delta p_T/p_T\simeq 0.4\% $ for $p_T\simeq 3 ~\rm GeV$ \cite{Dong:2019wtj}. 
Using the CDC measurements (both the transverse momentum 
and the angular information), we can 
compute the positron energy, which is then required 
to be equal to the electron energy 
(measured both in the ECL and CDC) in the c.m.\ frame. 
To suppress the backgrounds (especially those from neutrons), 
we further require a large missing energy 
such that the final state positron only deposits at most 
5\% of its energy in the ECL, 
and veto KLM activities including multi-hits or a cluster.

We next compute the event number of positrons.
Positrons at Belle II are mainly produced 
at the primary interaction point, 
via the Bhabha scattering process 
with the cross section
\cite{Peskin:1995ev}
\be
{d\sigma_B \over d \cos\theta^{*} } = {\pi \alpha^2 \over 2 s} 
{(3+\cos^2 \theta^{*})^2 \over 
(1+\cos\theta^{*})^2}, 
\label{eq:xsecBha}
\ee
where $\theta^{*}$ is the polar angle 
of the final state positron 
in the c.m.\ frame,
$s$ is the square of the center of mass energy, 
and $\alpha \simeq 1/137$ is the fine structure constant. 
Because SuperKEKB is an asymmetric 
collider, which collides
7 GeV electrons with 4 GeV positrons 
\cite{Belle-II:2018jsg}, 
the differential cross section 
of the Bhabha scattering in the lab frame is given by 
\be
{d\sigma_B \over d E } = {d\sigma_B \over d \cos\theta^{*} }
\frac{\sqrt{1-\beta^2}}{\beta E^{*}}, 
\label{eq:xsecBha:lab}
\ee
where 
$E$ is the energy of the final state positron 
in the lab frame, 
$\beta=3/11$, 
$E^*=\sqrt{s}/2$, and 
$\cos\theta^* = (\sqrt{1-\beta^2} E/E^*-1)/\beta$. 
In the lab frame, 
the energy of the final state positron $E$
is related to its polar angle $\theta$ via 
$E=E^* \sqrt{1-\beta^2}/(1-\beta \cos \theta)$. 
Thus, the minimum and maximum energy of the 
positron at the barrel region in the lab frame are 
$E_{\rm min}\simeq 4.35$ GeV 
(for $\theta = 128.7^\circ$) 
and $E_{\rm max}\simeq 6.62$ GeV 
(for $\theta = 32.2^\circ$) respectively.  
The total number of positrons in the barrel region is 
about $6\times 10^{11}$ 
with the total luminosity of $50 ~{\rm ab}^{-1}$.

\section{Standard Model Backgrounds}
\label{sec:BG}

SM backgrounds arise
when the SM particles that are produced in the 
collision between the final state positron 
and the ECL detector escape detection. 
Charged particles (such as electron and muon) 
are likely to be detected by 
the ECL and KLM detectors: 
The probability for positrons to penetrate 
the ECL is very small; 
the KLM detector,
which consists of an alternating sandwich of 4.7 cm thick iron plates and active detector elements \cite{Belle-II:2018jsg}, 
is very sensitive to the muon tracks, 
leading to negligible muon backgrounds via the KLM veto.\footnote{See appendix \ref{sec:chargedBG} for more detailed
  discussions on backgrounds due to charged particles.}
On the other hand, 
neutral particles (such as photon, neutron, and neutrino) have a significant 
probability to traverse 
the ECL and KLM detectors without being detected.
Backgrounds due to neutrinos are found to be 
negligible, due to the large $W/Z$ masses. 
Thus, the main backgrounds are due to 
photons and neutrons, which we discuss below.

\subsection{Photon-induced backgrounds}

We first discuss the photon-induced backgrounds.
Photon energy can be measured in the ECL detector, 
which is made up of CsI crystals with the length of 16 $X_0$ \cite{Shwartz:2017myl}, 
where $X_0=1.86 ~\rm cm$ 
is the radiation length of CsI \cite{ParticleDataGroup:2018ovx}.
The energy distribution of photons 
that are produced in the collision between 
a positron with energy $E$ and the ECL can be well approximated by 
\cite{Tsai:1966js}
\be
\frac{d N_\gamma}{d x_\gamma} (t,x_\gamma) \simeq \frac{1}{x_\gamma}\frac{(1-x_\gamma)^{(4/3)t}-e^{-(7/9)t}}{7/9+(4/3)\ln(1-x_\gamma)},
\ee
where $x_\gamma = E_\gamma/E$ 
with $E_\gamma$ being the energy of the photon, 
and $t X_0$ is the position of the photon 
in the ECL detector. 
Therefore, the probability of a photon 
carrying more than $95\%$
of the positron energy 
to escape the ECL detector is given by\footnote{The analytic method is able to yield very accurate results. For example, by
  using the NA64 parameters, we find that Eq.~\eqref{eq:gamma:BG} gives
  consistent results on the photon background with GEANT4 simulations in
  Ref.~\cite{Banerjee:2019pds}.}
\be
\int_{0.95}^1 dx_\gamma \frac{d N_\gamma}{d x_\gamma} (t=16,x_\gamma) \simeq 4.7 \times 10^{-8}. 
\label{eq:gamma:BG}
\ee 
This leads to $\sim$$2.8 \times 10^4$ potential background events after the 
ECL detector, for the $6\times 10^{11}$ positrons.
Although the probability for 
GeV-scale photons 
to penetrate the whole KLM 
(consisting of at least 60 cm iron plates in total \cite{Belle-II:2018jsg}) without producing  
KLM clusters is negligibly small, 
photons can also be 
absorbed by some non-instrumented setups 
(for example the magnet coil) between the ECL
and the KLM  \cite{Ferber}. 
For that reason the veto power of the KLM on photons 
is limited. 
To take into account such effects, 
we adopt the IFR veto efficiency at BABAR,
which is about $4.5\times 10^{-4}$
in the barrel region
\cite{BaBar:2008aby},
as the conservative   
estimate of the KLM veto efficiency, 
since the KLM veto efficiency is expected to be better than 
the IFR \cite{Corona}. 
\footnote{We note that the term ``veto efficiency'' 
represents the probability that the intended 
events do not satisfy the veto conditions. 
Thus, the ``veto efficiency'' actually describes the 
inefficiency of the veto. 
In other words, a smaller value of the veto efficiency means  
a better veto (on the intended events).}
This then leads to $13$ background events due to photons, 
for the $6\times 10^{11}$ 
positrons. 
We note that a small fraction of secondary photons can  
leak into the endcap region, 
where the veto efficiency is somewhat reduced;  
in our analysis, we have neglected this small effect.
We also note that if the KLM veto efficiency 
can be improved by one order of magnitude as 
compared  to the IFR,
the photon backgrounds will 
decrease to be about a single event. 
On the other hand, 
if the KLM veto efficiency turns out to be inferior to 
that of the IFR, 
an increase in photon-induced background events is expected. 
To account for this possibility, 
in section \ref{sec:results}, we 
compute the Belle II sensitivity both with the 
background events analyzed in our current analysis 
and with a more conservative background level. 
The sensitivities under these different 
background assumptions are given in Fig.~(\ref{fig:BelleIIDP}).

\subsection{Neutron-induced backgrounds}

We next discuss the neutron-induced backgrounds.
Neutrons with energy of several GeV are mainly 
produced by photo-nuclear reactions between the 
positron and the ECL detector \cite{ParticleDataGroup:2018ovx}.
To estimate such backgrounds, 
we simulate collisions 
of $10^9$ positrons 
with $4.35$ GeV energy onto a CsI target with one $X_0$,
by using GEANT4 (version 11.0) \cite{GEANT4:2002zbu} 
with the \texttt{FTFP\_BERT} physics list. 
Our choice of the positron energy in the simulation is 
motivated by the fact that 
$\sim 50\%$ of the positrons are in the 
first tenth of the entire energy range in the barrel region, 
according to Eq.~\eqref{eq:xsecBha:lab}.
We only simulate a fraction of positrons with a thin CsI target (with one $X_0$)
because the full simulation with a 16-$X_0$ CsI target is extremely time-consuming. 
The simulation results with the thin CsI target 
are acceptable for our purpose, 
because neutrons with significant energy 
are mainly produced within the first radiation length, 
which are confirmed in our simulations 
with a 2-$X_0$ CsI target.

To ensure
that the missing energy 
is mainly caused by neutrons, 
we only select the GEANT4 simulated events 
that contain at least one neutron with energy  
exceeding 3 GeV. There  
are 4950 events in the $10^9$ simulations that
satisfy this selection cut. 
We then compute the total energy deposited 
in the ECL, by taking into account 
both the deposited energy in the first $X_0$ calculated by GEANT4, 
and the kinetic energy of $e^\pm$ and $\gamma$. 
We further include the kinetic energy of protons 
with momentum less than 0.6 GeV, 
because such protons have
a gyroradius radius $\lesssim 1.3$ m 
in the ECL where $B=1.5$ T \cite{Belle-II:2018jsg}, 
and thus can deposit the kinetic energy when orbiting around. 
We do not add the kinetic energy of $\pi^\pm$ to the deposited energy, 
because $\pi^\pm$ decays primarily 
into a neutrino and a muon which deposit 
negligible energy to the ECL. 
We then require the deposited energy in the ECL
to be less than 5\% of the energy of the positron; 
there are 100 events after this detector cut. 
We further veto events that  
contain protons or $\pi^\pm$
with momentum exceeding 0.6 GeV, 
because 
these charged hadrons can either deposit significant energy 
in the ECL and/or produce tracks in the KLM.
There are 64 events after this veto.

Next we classify the remaining events 
according to the number of neutrons 
that have kinetic energy exceeding 280 MeV, 
the energy threshold for 
hadronic showers \cite{Cerrito:2017tiy}. 
There are 13 events with a single neutron 
and 51 events with more than 2 neutrons. 
We compute the probability for a neutron to penetrate 
a target with length $L$ via 
\cite{Grupen:2008zz, Andreas:2013lya} 
\be
P = \exp (-L/\lambda_0), 
\label{eq:neutron:prob}
\ee 
where $\lambda_0$ is the hadronic interaction length.
The KLM has $\sim 3.9 \lambda_0$, 
and the ECL has $\sim 0.8 \lambda_0$ \cite{Belle-II:2018jsg}. 
We note that Eq.~\eqref{eq:neutron:prob} can yield consistent 
results on neutron-induced backgrounds as compared with GEANT4 simulations 
in the context of the NA64 experiment \cite{Andreas:2013lya}.
Thus, the probability for a neutron to penetrate the remaining $15$ $X_0$'s of the ECL
and the KLM is $P\simeq 0.01$.\footnote{ See e.g., Refs.~\cite{Belle:2015qal, Aushev:2014spa} for more detailed
  analyses on the role of the KLM detector as a hadronic veto.}
Rescaling this to the $6\times 10^{11}$ positrons,
one expects $\sim$81 background 
events due to neutrons in total.

We note that there is another source of neutrons 
from the beam backgrounds (dominated by 
10-100 keV neutrons), 
which can also produce KLM hits \cite{Belle-II:2010dht}  
and thus complicates the situation.
Because of the beam backgrounds, 
one cannot veto events with any hits in the KLM \cite{Ferber}. 
Fortunately, unlike neutrons with kinetic energy above 280 MeV 
which are expected to produce 
multi-hits or a cluster in the KLM, 
a single beam background neutron is usually 
absorbed in one scintillator strip \cite{Aushev:2014spa}. 
For that reason, in our analysis, 
we only select neutrons above 
280 MeV, which can be well controlled 
by the veto on multi-hits or a cluster in the KLM.
However, 
since there is already a neutron with energy above 3 GeV 
in our selected events, including another neutron below 
280 MeV would further suppress the background, 
leading to an even smaller neutron background.
Thus, our analysis  
serves as a conservative estimate of the neutron backgrounds. 

Taking into account backgrounds 
from both neutrons and photons, one 
expects at most $\sim$94 background events for the 
$6\times 10^{11}$ positrons. 

\section{Sensitivity on dark matter in dark photon models}
\label{sec:results}

To show the sensitivity of 
the ``disappearing positron track'' channel
on DM, we consider  
a new physics model in which 
DM interacts with the SM through a DP 
\cite{Jaeckel:2010ni,Fabbrichesi:2020wbt}. 
In new physics scenarios where 
the SM gauge group is extended by an additional $U(1)$, 
DP can naturally arise, 
either via the kinetic mixing portal 
\cite{Holdom:1985ag,Foot:1991kb}, 
or via the Stueckelberg mass mixing portal 
\cite{Kors:2005uz,Feldman:2006ce, Feldman:2006wb, Cheung:2007ut, Feldman:2007wj, Du:2019mlc,Du:2021cmt}.  
For both portals, field redefinitions are necessary to recast the kinetic terms into canonical forms and to diagonalize the mass matrix for the neutral gauge bosons. These processes lead to DP interactions with matter fields 
both in the SM sector and in the dark sector. 
In the case of a small mixing parameter, 
the interaction Lagrangian can be parameterized as follows 
\cite{Feldman:2007wj,Fabbrichesi:2020wbt}
\begin{equation}
    {\cal L}_{\rm int} = A'_\mu 
    (e Q_f \epsilon \bar f \gamma^\mu f 
    + g_\chi \bar \chi \gamma^\mu \chi), 
\end{equation}
where $A'_\mu$ is the DP  
with mass $m_{A'}$, 
$\chi$ is the Dirac DM with mass $m_\chi$, 
$f$ denotes the SM fermion with electric charge $Q_f$, 
$\epsilon$ is the small mixing parameter,
$e$ is the QED coupling constant, 
and $g_\chi$ is the hidden coupling constant.
In our analysis we fix $m_\chi=m_{A'}/3$ such that 
$A'$ decays dominantly into DM 
in the parameter space of interest where 
$g_\chi \gg e \epsilon$.

We compute the  
signal events from both diagrams shown in 
Fig.~(\ref{fig:Feynmandig}). 
We first compute the
annihilation process $e^+ e^-_{A} \to A' \to \chi \bar{\chi}$; 
the cross section is given by
\be
\sigma_{\rm ann}(\sqrt{s})=\frac{e^2 \epsilon^2 \alpha_D}{3}
\frac{s+2m_\chi^2}{(s-m_{A^\prime}^2)^2+\Gamma_{A'}^2 m_{A'}^2}
\sqrt{1-\frac{4 m_\chi^2}{s}},
\label{eq:DPxsec}
\ee
where $\alpha_D=g_\chi^2/4\pi$, 
$\Gamma_{A'}$ is the decay width of the DP,  
and $s = 2m_eE'+2m_e^2 = 2 m_e E_{A'}$  
with $E'$ being the energy of the positron 
at the collision point
and $E_{A'}=E'+m_e$ being the energy of $A'$. 
Note that we have $E' \leq E$ where $E$ 
is the positron energy before entering ECL.
The partial decay width of the DP into DM is 
\be
\Gamma (A' \to \bar \chi \chi)=\frac{m_{A'}\alpha_D}{3}
\left(1+2\frac{m_\chi^2}{m_{A'}^2} \right)\sqrt{1-\frac{4m_\chi^2}{m_{A'}^2}}.
\label{eq:DPdecay}
\ee
Because the invisible decay width is much larger than 
the visible ones in the parameter space of interest, 
we use $\Gamma_{A'} \simeq \Gamma (A' \to \bar \chi \chi)$ in our analysis. 
The signal events in the annihilation process 
can be computed by \cite{ Andreev:2021fzd,Gninenko:2018ter,Marsicano:2018krp}  
\be
N_{\rm ann} = \mathcal{L}  \int_{E_{\rm min}}^{E_{\rm max}} dE\frac{d\sigma_{B}}{dE}  \int_{0.95E}^{E+m_e} dE_{A'}   
n_e  T_e(E'=E_{A'}-m_e,E,L_T) \sigma_{\rm ann}(E_{A'}) ,
\label{eq:NDPann}
\ee
where $\mathcal{L}=50$ ab$^{-1}$ is the integrated luminosity, 
$n_e$ is the number density of the electron in CsI, 
and ${d\sigma_{B}}/{dE}$ is given in Eq.~\eqref{eq:xsecBha:lab}.
Here $T_e(E',E,L_T)$ 
is the positron differential track-length distribution \cite{Bjorken:1988as,Marsicano:2018krp,Marsicano:2018glj,Tsai:1966js} 
where $L_T=16 X_0$ is the thickness of the ECL target. 
The expression of $T_e$ is given in 
appendix \ref{sec:TrackLength}. 
The integration of $E_{A'}$ is 
performed for $E_{A'}>0.95E$ so that the 
positron deposits less than 5\% of its 
original energy in the ECL.

We next compute the bremsstrahlung process. 
In the parameter space of interest, 
the signal is dominated by the on-shell produced $A'$.
Thus, the signal events are given by
\bea
N_{\rm bre}=\mathcal{L}  \int_{E_{\rm min}}^{E_{\rm max}} dE\frac{d\sigma_{B}}{dE}  \int_{0.95E}^{E-m_e} dE_{A'} 
\int_{E_{A'}}^{E} d E' \, 
n_N  T_e(E',E,X_0)
{d\sigma_{\rm bre} \over dE_{A'}},
\label{eq:NDPbrem} 
\eea
where $n_N$ is the number density of I (or Cs).
Here $d\sigma_{\rm bre}/dE_{A'}$ are 
the differential cross section of the 
on-shell produced $A'$ 
\cite{Bjorken:2009mm, Gninenko:2017yus,Liu:2016mqv, Liu:2017htz},
\be
{d\sigma_{\rm bre} \over d E_{A'}}= (\phi_I + \phi_{\rm Cs})
\frac{4\alpha^3\epsilon^2}{E'}  \frac{x(1-x+x^2/3) }{m^2_{A'} (1-x)+m_e^2x^2},
\ee
where $x \equiv {E_{A'}}/{E'}$, 
and $\phi_N$ denotes the effective flux of photons 
from nucleus $N$ \cite{Bjorken:2009mm}:
\be
\phi_N 
=\int_{t_{\rm min}}^{t_{\rm max}} dt \, \frac{t-t_{\rm min}}{t^2} 
 \Bigg[ {Z a^2 t \over (1+ a^2 t) (1+ t/d)} \Bigg]^2,
\ee
with  $t_{\rm min}=(m_{A'}^2/2E')^2$,
$t_{\rm max}=m_{A'}^2+m_e^2$,
$a= 111 m_e^{-1} Z^{-1/3}$, and
$d=0.164 A^{-2/3}$ GeV$^2$. 
We use $Z=53 \, (55)$ and $A =127 \, (133)$ for I (Cs).
Here we only consider 
the dominant elastic form factor.

\begin{figure*}[htbp]
\begin{centering}
\includegraphics[width=0.45 \textwidth]{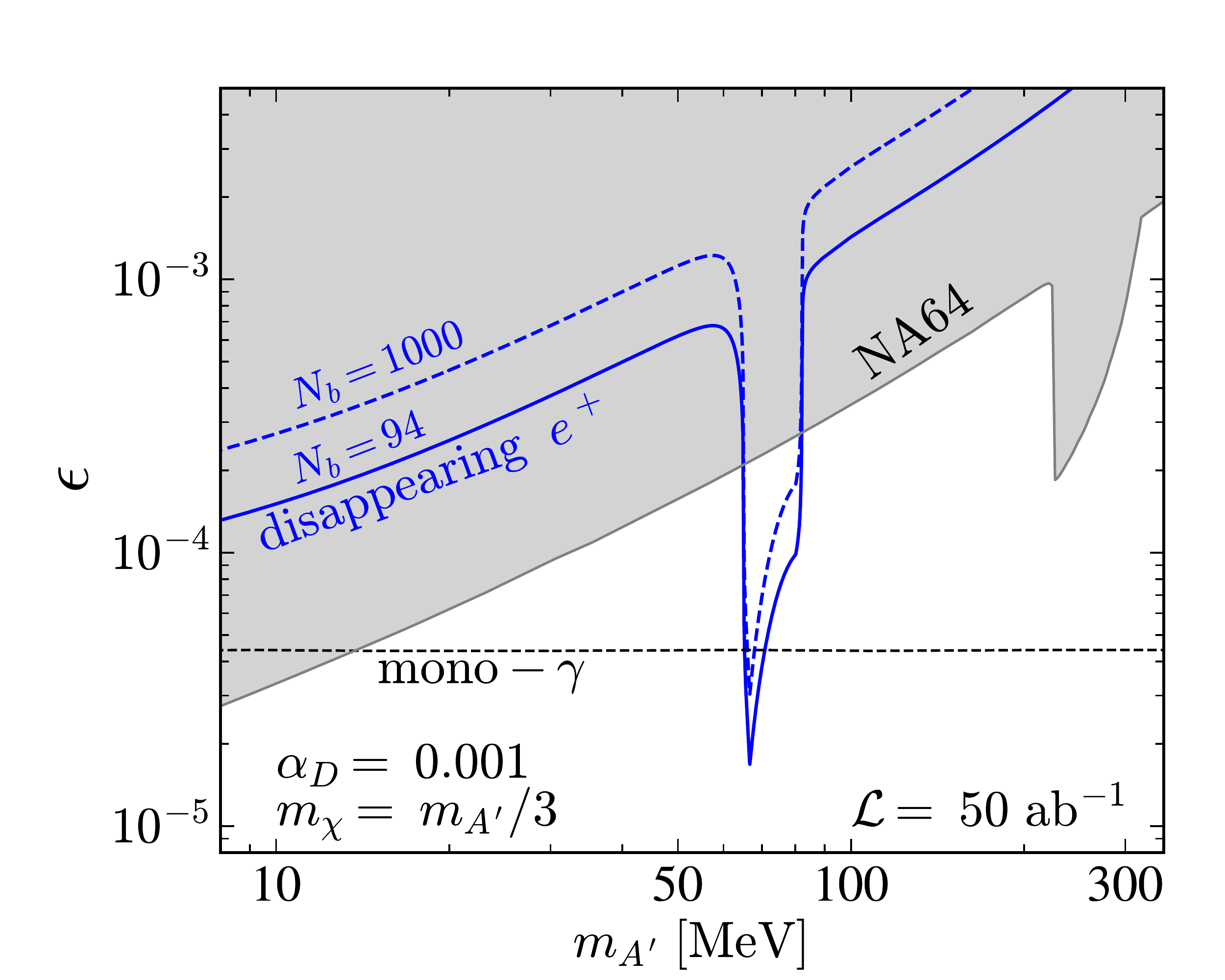}
\caption{Belle II sensitivities on 
the coupling parameter $\epsilon$ 
between the DP and SM particles
with $50 ~{\rm ab}^{-1}$ 
integrated luminosity 
as a function of the DP mass $m_{A'}$
from the ``disappearing positron track'' 
with $N_b=94$ (blue solid) and with $N_b=1000$ (blue dashed). 
Here we fix $\alpha_D=0.001$ and $m_\chi=m_{A'}/3$. 
Also shown are the 
Belle II sensitivity from the mono-photon channel 
with $50 ~{\rm ab}^{-1}$ 
(black dashed), 
and the NA64 constraints (gray shaded region); 
see main text for details. 
}
\label{fig:BelleIIDP}
\end{centering}
\end{figure*}

Fig.~(\ref{fig:BelleIIDP}) shows 
the expected 90\% CL limits on 
the dark photon coupling parameter 
$\epsilon$ 
from the  ``disappearing positron track'' channel, 
as a function of the DP mass, 
where we take $m_{A'}=3 m_\chi$, 
$\mathcal{L}=50~{\rm ab}^{-1}$, 
and $\alpha_D = 0.001$.
We compute the 90\% CL limits 
by using  
$N_{s}/\sqrt{N_b}=\sqrt{2.71}$ \cite{Liang:2021kgw} 
where $N_{s}=N_{\rm ann} + N_{\rm bre}$ 
and $N_b = {94}$.
We find that in the narrow mass window, 
${66~\rm MeV}\lesssim m_{A'} \lesssim 82 ~\rm MeV$, 
the annihilation process with the atomic electrons dominates; 
outside this region, the bremsstrahlung process dominates. 
The narrow mass window can be explained by the 
Breit-Wigner resonance in Eq.~\eqref{eq:DPxsec}, where 
one has  
$m_{A'} \simeq \sqrt{s} \simeq \sqrt{2 E_{e^+} m_e}$.   
The positron energy in the barrel region of the ECL 
is in the range of 4.35-6.62 GeV, which leads to 
${66~\rm MeV}\lesssim m_{A'} \lesssim {82 ~\rm MeV}$ 
if the $<5\%$ energy difference between 
secondary positrons and the incident position is neglected.
We thus refer to the narrow mass window where 
the annihilation process can proceed via the 
Breit-Wigner resonance of the mediator 
as the resonance region. 
The Belle II sensitivity in the resonance is significantly 
enhanced, as shown in Fig.~(\ref{fig:BelleIIDP}).
\footnote{We also anticipate such an enhancement for other 
mediators. In appendix \ref{sec:scalar}, we consider an alternative 
model where DM interacts with the SM sector via a scalar mediator. 
We find that the Belle II sensitivity on the 
scalar mediator model is similar to that of the 
dark photon model.}

The DP models can also be 
searched for in the mono-photon channel 
$e^+ e^- \to \gamma A'$ 
at Belle II 
\cite{Belle-II:2018jsg}
and by the missing energy signature at NA64 
\cite{Andreev:2021fzd,Banerjee:2019pds}. 
The Belle II sensitivity from the mono-photon channel 
with $50 ~{\rm ab}^{-1}$, shown in Fig.~(\ref{fig:BelleIIDP}),
is rescaled from the result with $20~\rm fb^{-1}$ in Ref.~\cite{Belle-II:2018jsg}, 
assuming that the limit on $\epsilon$ is proportional to $\mathcal{L}^{-1/4}$.
\footnote{Note that such a naive rescaling may not hold  
for the dark photon mass below 2 GeV, where the  
cosmic ray background becomes important and its 
characterization is yet to be completed \cite{Belle-II:2022cgf}. 
Nevertheless, we provide such a rescaling of the mono-photon sensitivity 
as a comparison to our newly proposed channel.}
Fig.~(\ref{fig:BelleIIDP}) shows the 
 NA64 constraints with $2.84\times 10^{11}$ electrons on target 
 in the $\alpha_D=0.001$ case.
The NA64 constraints away from the resonance region, 
which is $\sim(200-300)$ MeV, 
are from Ref.~\cite{Banerjee:2019pds}.
We compute the NA64 constraints in 
 the resonance region 
 by taking into account both  
 the annihilation process 
and the bremsstrahlung process; 
the detailed calculations are given 
in 
appendix \ref{sec:NA64calc}, 
where we also carry out a comparison between 
our analytic method and the results 
given in Ref.~\cite{Andreev:2021fzd}, to demonstrate  
the accuracy of 
our calculation.
For the $\alpha_D =0.001$ case, 
if we take $N_b=94$,
we find that 
the best limit on $\epsilon$ from 
the new ``disappearing positron track'' channel 
is $\epsilon \lesssim 1.7\times 10^{-5}$,
which occurs in the vicinity of $m_{A'} \simeq 66$ MeV, 
surpassing both the mono-photon channel at Belle II 
and the NA64 constraints. 
As $\alpha_D$ increases, 
the limits on $\epsilon$ are slightly weakened;
for example, the best limit is  
$\epsilon \lesssim 2.6 \times 10^{-5}$
for the case of $\alpha_D =0.1$.
See appendix \ref{sec:aDdepend} 
for the detailed dependence of the limits on $\alpha_D$.

In our benchmark model point, we have used the mass relation 
$m_{A'} = 3 m_\chi$. 
We note that the limits have a weak dependence on 
$m_\chi$, if it is decreased to even smaller values. 
For example, 
in the $m_{A'} =66~\rm MeV$ case, 
the number of events is only 
decreased by $\sim0.2\%$, if we change 
$m_{A'} = 3m_\chi$ to $m_{A'} = 5m_\chi$. 
This is because for the invisible dark photon model 
the signal process can be well approximated by 
the production of an on-shell dark photon.

We note that, despite our efforts to reproduce the Belle II
conditions to the best of our ability, 
the photon- and neutron-induced backgrounds estimated 
in our analysis could still be subject to large uncertainties. 
To illustrate the impacts of these potential uncertainties on the 
sensitivity, we also compute the sensitivity with a background that is 
about one order of magnitude larger than our initial analysis. 
This is shown as the blue dashed curve with $N_b=1000$ 
in Fig.~(\ref{fig:BelleIIDP}). 
We find that the sensitivity on $\epsilon$ 
is weakened by a factor of $\simeq 1.8$, 
if the background is increased from $N_b=94$ to $N_b=1000$. 
This is because the new physics signal is proportional to $\epsilon^2$, 
while the expected limit on the new physics signal is proportional to $\sqrt{N_b}$.  
Even with $N_b=1000$, we find that  
the sensitivity from the ``disappearing positron track'' in the resonance region, 
${66~\rm MeV}\lesssim m_{A'} \lesssim 82 ~\rm MeV$, 
remains stronger than the NA64 constraint 
and can still be comparable to the mono-photon sensitivity from Belle II.

Here we discuss the possibility of using 
the ``disappearing electron track'' as a control sample. 
In this control sample, 
the electron in the final state of the Bhabha scattering 
interacts with the ECL to produce DM, 
with the positron being fully reconstructed. 
The control sample should yield a similar signal in the 
bremsstrahlung process 
as the ``disappearing positron track'', 
but its signal in the annihilation process is very small. 
One might expect a null result in the annihilation process for  
the ``disappearing electron track'' because 
there are no atomic positrons in the ECL.  
However, secondary positrons can be generated when an  
electron traverses the ECL. 
By comparing the differential track lengths of positrons 
for incident electrons and positrons, 
which are given in Appendix \ref{sec:TrackLength}, 
we find that the secondary positrons 
that carry $>95\%$ energy of the incident electron  
are smaller than those originating from an incident positron 
by a factor of $\sim 5\times 10^{-4}$. 
Thus, the annihilation process in the control sample 
is expected to be smaller than the ``disappearing positron track'' 
by the same reduction factor. 
Because backgrounds are expected to be the same for both
the ``disappearing electron track'' and the ``disappearing positron track'' channels, 
one can use the control sample to cross check the background analysis.  
The most interesting parameter space in our analysis occurs at 
$m_{A'} \simeq 66~\rm MeV$ and $2\times 10^{-5} \lesssim \epsilon \lesssim 10^{-4}$, where the 
sensitivity is dominated by the annihilation between positrons 
and atomic electrons in the ECL. 
If excess events beyond the SM backgrounds were to be observed in this region 
for the ``disappearing positron track'', 
one could cross check the signal with the control sample, 
which is expected to yield a signal that is 
a factor of $\sim 5\times 10^{-4}$ smaller.

\section{Conclusions}
\label{sec:conclusion}

In this paper,
we propose a new ``disappearing positron track'' channel 
at Belle II to search for DM, 
where DM are generated via collisions between 
positrons and the ECL.
The major backgrounds are
due to photons and neutrons produced in the 
same collisions.
We design a set of detector cuts 
to reconstruct such 
a new signal from the Belle II data, 
as well as to suppress various SM backgrounds. 
We compute the sensitivity of the new channel on 
the invisible dark photon model.
We find 
that the new channel at Belle II 
can probe 
the dark photon coupling parameter 
$\epsilon \simeq 1.7\times 10^{-5}$
with 50 ab$^{-1}$ data 
for dark photon mass at $\sim$66.66 MeV, 
surpassing both the mono-photon channel at Belle II 
and the missing energy channel at NA64. 
We note that the disappearing track signal on positrons 
can be 
further extended to other SM particles 
at different particle colliders, 
thus presenting an opportunity to 
probe various new physics models with diverse interactions.

\begin{acknowledgments}

We thank Luigi Corona, 
Torben Ferber, 
and Li-Gang Xia
for correspondence and discussions. 
We also would like to thank the anonymous referee for many valuable comments.
The work is supported in part by the 
National Natural Science Foundation of China under Grant Nos.\ 12275128 
and 12147103, by the 
Science and Technology Program of Guangzhou No.\ 2019050001,
and by the
Guangdong Major Project of Basic and Applied Basic Research 
No.\ 2020B0301030008.
\end{acknowledgments}

\appendix

\section{Track-length distribution of positrons and electrons}
\label{sec:TrackLength}

For an incident positron with an initial energy $E$ to 
enter a target with thickness $L_T$, 
the differential track-length distribution of positrons
as a function of the positron energy $E'$
can be computed by \cite{Marsicano:2018krp,Marsicano:2018glj}
\be
T_e(E',E,L_T)=X_0\int_{0}^{L_T/X_0} I_e(E',E,t) dt
\label{eq:tracklength},
\ee
where $X_0$ is the radiation length of the target. 
Here $I_e(E',E,t)$ is the energy distribution of $E'$  
at the depth $t X_0$, 
which can be computed iteratively 
such that 
$I_e= \sum_i I_e^{(i)}$ 
where $I_e^{(i)}$ denotes 
the $i$-th generation positrons \cite{Tsai:1966js}.
We adopt the analytical 
model of Ref.~\cite{Tsai:1966js} up to 
second-generation positrons, 
which are found to be 
in good agreement with simulations 
in Ref.~\cite{Marsicano:2018krp}.
The contributions from the first two generations 
are \cite{Tsai:1966js}
\bea
\label{eq:tracklength:g1}
I^{(1)}_e(E',E,t)&=&\frac{1}{E}
\frac{(\ln (1/v))^{b_1 t-1}}{\Gamma( b_1 t)}, \\
I^{(2)}_e(E',E,t)&=&\frac{2}{E}
\int_{v}^{1} \frac{d x}{ x^2} \frac{1}{b_2+b_1\ln(1-x)} 
 \nonumber \\ &&
 \label{eq:tracklength:g2}
 \times \left[ \frac{(1-x)^{ b_1 t}-(1-v/x)^{b_1 t}}{b_1 \ln  \left[ (x-x^2)/(x-v)  \right] } \right. 
+ \left.
  \frac{e^{- b_2 t} -(1-v/x)^{b_1 t}}{b_2 +b_1 \ln (1-v/x)}
\right],
\eea
where $b_1=4/3$, $b_2=7/9$, $v=E'/E$. 
We note that the 
analytical expression for the positron track length (with the first two generations) can 
yield consistent results with GEANT4 simulations; 
see e.g., figure 4 of Ref.~\cite{Marsicano:2018krp} 
for the comparison in the case of 
an aluminum target.

Note that Eqs.~(\ref{eq:tracklength}, \ref{eq:tracklength:g1}, \ref{eq:tracklength:g2})
can also be used to compute 
the track length distribution of electrons
with an incident electron. 
We compute 
the track length distribution 
of positrons (electrons) 
with an incident electron (positron) via 
\be
\bar{T}_e(E',E,L_T)=X_0\int_{0}^{L_T/X_0} I_e^{(2)}(E',E,t) dt
\label{eq:tracklength-dif}, 
\ee
where we have only used $I_e^{(2)}$. 
This is because 
$I_e^{(1)}$ describes the effects of bremsstrahlung, 
and $I_e^{(2)}$ subsequently describes the production 
of the electron and positron pair due to 
the emitted photon \cite{Tsai:1966js}.

\section{Dependence of signal on $\alpha_D$}
\label{sec:aDdepend}

Here we investigate the dependence of the dark photon constraint from the ``disappearing positron track'' channel at Belle II on the 
gauge coupling constant $\alpha_D$ in the hidden sector.

Dark photons can be produced either in the bremsstrahlung process or in the annihilation process. 
In the bremsstrahlung process, because dark photons are on-shell produced, the cross section does not depend on $\alpha_D$. 
However, in the annihilation process, the cross section 
depends on $\alpha_D$ through the 
Breit-Wigner form of the dark photon propagator, 
since the hidden decay width of the dark photon 
is proportional to $\alpha_D$.
\begin{figure}[tbhp]
\begin{centering}
\includegraphics[width=0.5 \textwidth]{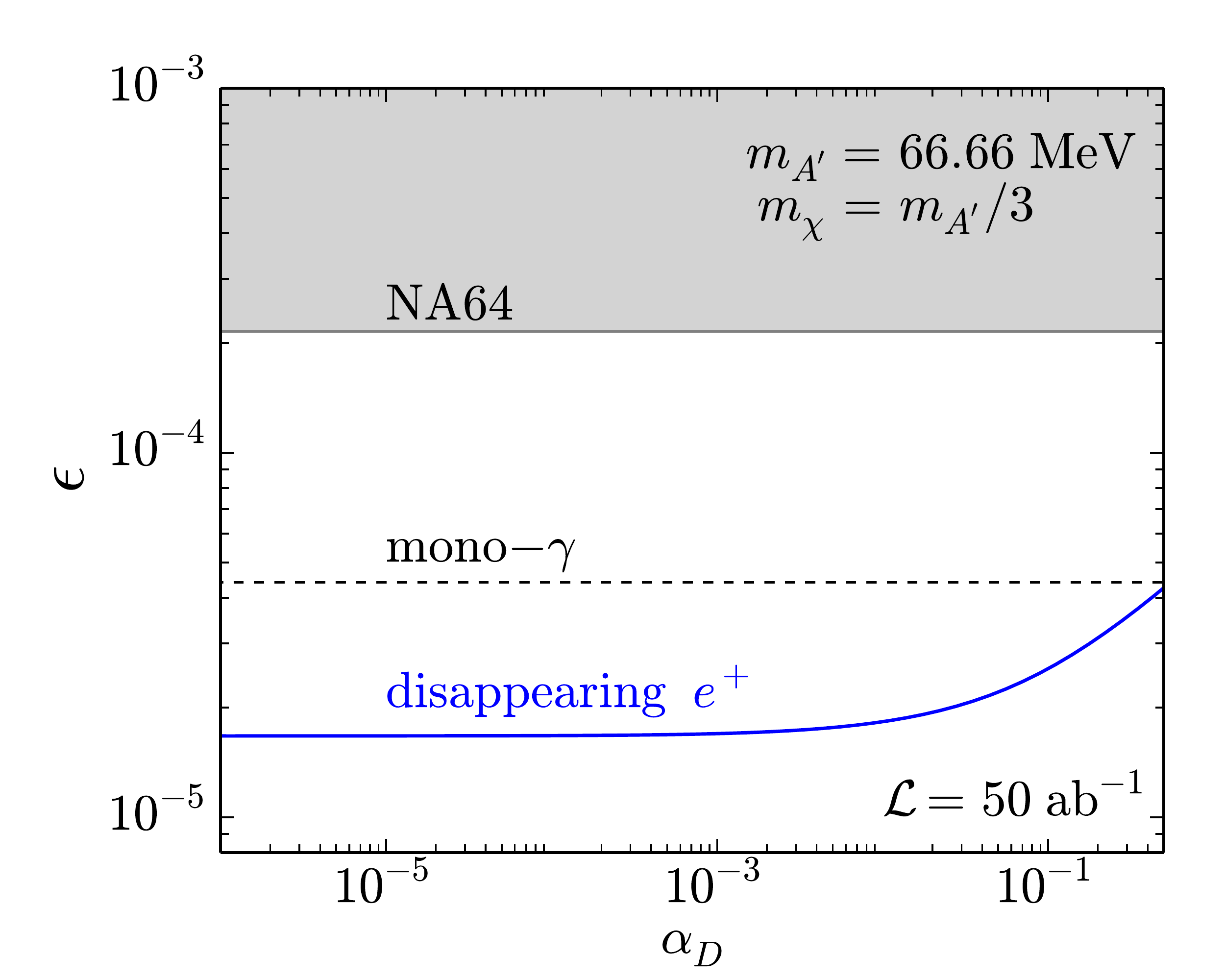}
\caption{Belle II sensitivity on 
the coupling parameter $\epsilon$ with $50 ~{\rm ab}^{-1}$ 
integrated luminosity 
as a function of $\alpha_D$
from the ``disappearing positron track'' (blue solid),
where $m_{A'}=66.66 ~\rm MeV$ and $m_\chi=m_{A'}/3$. The Belle II sensitivity from the mono-photon channel 
with $50 ~{\rm ab}^{-1}$ 
(black dashed) 
is rescaled from the result with $20~\rm fb^{-1}$ in Ref.~\cite{Belle-II:2018jsg}.
The NA64 constraint with $2.84\times 10^{11}$ electrons on target (gray shaded region) is taken from Ref.~\cite{Banerjee:2019pds}. 
} 
\label{fig:BelleIIDP_aDeps}
\end{centering}
\end{figure}
Fig.\ (\ref{fig:BelleIIDP_aDeps}) shows 
the constraints on the coupling parameter $\epsilon$ 
as a function of $\alpha_D$,
in which $m_{A'}$ is fixed to be $66.66 ~\rm MeV$. 
The upper bound on $\epsilon$ decreases with  $\alpha_D$ 
until it saturates at $\epsilon \lesssim 1.7 \times 10^{-5}$ for $\alpha_D \lesssim 0.001$. 
Constraints from the missing energy signature at NA64 and from the mono-photon channel at Belle II
are also shown in Fig.\ (\ref{fig:BelleIIDP_aDeps}), which are 
independent of $\alpha_D$.
It is remarkable that the ``disappearing positron track'' channel for $\alpha_D \lesssim 0.5$
leads to a better constraint 
than the mono-photon channel at Belle II and the missing energy signature at NA64.

\section{NA64 constraints}
\label{sec:NA64calc}

In this section, we compute  
the NA64 constraints. 
Note that Ref.~\cite{Andreev:2021fzd} provided 
the NA64 constraints on dark photon with $\alpha_D =0.1$ and $\alpha_D =0.5$.
In our calculation,
we compute the NA64 constraints in 
 the resonance region 
 by taking into account the number of signal events both from the 
bremsstrahlung process ($N_{\rm bre}$) and from the annihilation process ($N_{\rm ann}$).

The NA64 constraint on the coupling parameter $\epsilon$ taking into account only the bremsstrahlung process (denoted as 
$\epsilon_{\rm bre}$) has been obtained in Ref.~\cite{Banerjee:2019pds, NA64:2021xzo},
by requiring $N_s =2.3$ where $N_s$ 
is the number of signal events.
Since the signal from the bremsstrahlung process is proportional to 
$\epsilon^2$ and independent on $\alpha_D$, we compute the number of signal 
events from the bremsstrahlung process via
\bea
N_{\rm bre}=N_s \left(\frac{\epsilon^2}{\epsilon_{\rm bre}^2}\right), 
\eea
where $N_s=2.3$.

We compute the number of signal events from the annihilation process via
\bea
N_{\rm ann}&=&\epsilon_d N_{\rm EOT}    \int_{0.5E}^{E+m_e} dE_{A'} \sigma_{\rm ann}(E_{A'})  
n_e T_e(E_{A'}-m_e,E,L_T),
\label{eq:NDPann-NA64}
\eea
where $E=100$ GeV is the energy of the incident electrons, 
$N_{\rm EOT}= 2.84\times 10^{11}$ is the number of electrons on target, 
$n_e$ is the electron number density of the lead target, 
$\epsilon_d\simeq 0.5$ is the detection efficiency \cite{Banerjee:2019pds}, 
$\sigma_{\rm ann}$ is the cross section of the annihilation process 
$e^- e^+ \to \chi \bar{\chi}$, $E_{A'}$ is the energy of the dark photon, $L_T$ is 
the length of the target, and $T_e(E'=E_{A'}-m_e,E,L_T)$ is the positron differential track-length distribution.

\begin{figure}[tbhp]
\begin{centering}
\includegraphics[width=0.5 \textwidth]{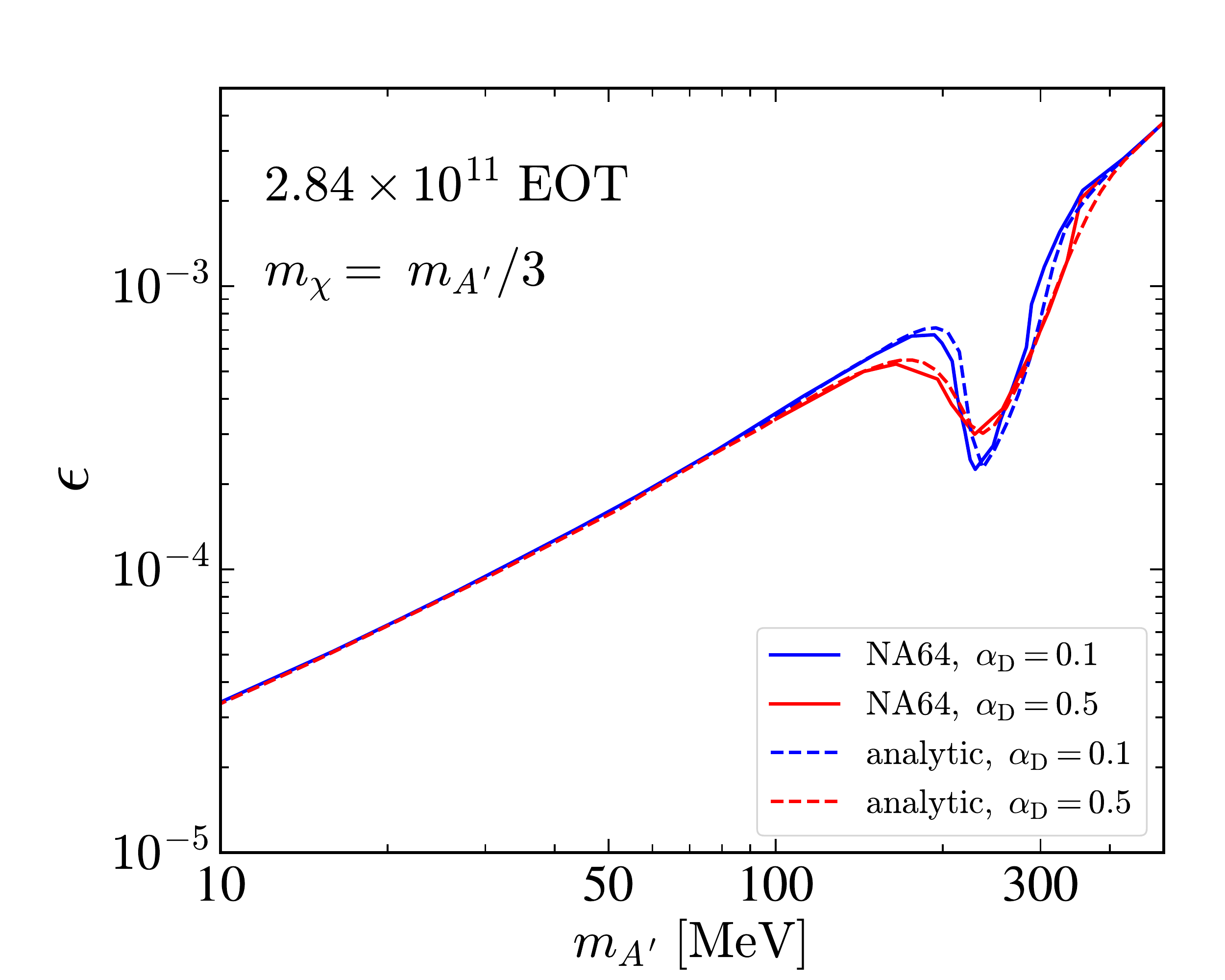}
\caption{NA64 constraints 
(with $2.84 \times 10^{11}$ electrons on target) 
obtained in our analytical calculations (dashed lines) and in Ref.~\cite{Andreev:2021fzd} 
where GEANT4-based simulations are employed (solid lines). 
The comparison is carried out for the mass relation 
of $m_\chi =m_{A'}/3$ 
and for two different couplings:  
$\alpha_D =0.1$ (blue) and $\alpha_D =0.5$ (red).} 
\label{fig:NA64DP}
\end{centering}
\end{figure}

We obtain the 90\% C.L. constraints by using the criterion $N_{\rm ann}+N_{\rm bre}=2.3$. 
We recompute the NA64 limits in Fig.~(\ref{fig:NA64DP}) 
and compare them with those presented in the 
NA64 analysis \cite{Andreev:2021fzd},  
where GEANT4-based simulations are employed. 
As shown in Fig.~(\ref{fig:NA64DP}),
our calculation yields consistent results with Ref.~\cite{Andreev:2021fzd}. 
In particular, 
in the vicinity of $m_{A'} \simeq 250$ MeV, 
where the $e^+e^-$ annihilation process dominates, 
our analytic method utilizing Eq.~\eqref{eq:NDPann-NA64} 
successfully reproduces 
the NA64 results \cite{Andreev:2021fzd}, 
for both $\alpha_D=0.5$ and $\alpha_D=0.1$.
This demonstrates the accuracy of our method, which utilizes the 
analytic expressions of the positron track length distribution 
(with the first two generations), 
in predicting dark photon signal events 
in the region 
dominated by the $e^+e^-$ annihilation process.

\section{SM backgrounds due to charged particles}
\label{sec:chargedBG}

We have assumed in the analysis 
that charged particles lead to 
insubstantial backgrounds compared to neutral particles, 
such as photons and neutrons. 
Here, we provide a more detailed discussion of potential background sources due to charged particles.

Backgrounds can arise from pairs of charged particles generated in $e^+e^-$ collisions at the primary vertex, 
with notable examples including $\pi^+ \pi^-$ and $\mu^+ \mu^-$.
To be classified as a background contribution, 
the charged pair must satisfy the following criteria: 
The negatively charged particle within the pair must 
effectively mimic a fully reconstructed $e^-$, 
exhibiting successful reconstructions in both CDC and ECL (with nearly all of its energy 
deposited in ECL).
Additionally, the positively charged particle 
should emulate a poorly reconstructed $e^+$, 
featuring a robust reconstruction in CDC, 
but with less than 5\% of its energy detected by ECL 
and no significant energy deposition in KLM (to avoid KLM veto).

We first discuss backgrounds due to the $\pi^+\pi^-$ final state 
generated in $e^+e^-$ collisions at the primary vertex. 
First note that $\sigma(e^+e^- \to \pi^+\pi^-)$ at Belle II 
is significantly suppressed by the pion form factor, 
making it substantially smaller than leptonic final states, such as $\mu^+\mu^-$. 
We estimate $\sigma(e^+e^- \to \pi^+\pi^-)$  
by utilizing the measurement of $d\sigma(\pi^+\pi^-)/dz$ in $e^+e^-$ collisions 
conducted in Ref.~\cite{Belle:2013lfg},  
where $z= 2 E_\pi /\sqrt{s}$ with $\sqrt{s} = 10.52 ~\rm GeV$. 
To mimic the $e^+e^-$ final state, $z$ has to be close to 1. 
However, due to limitations of the method used in Ref.~\cite{Belle:2013lfg}, 
$d\sigma(\pi^+\pi^-)/dz$ is provided only for $z<0.98$. 
Since $d\sigma(\pi^+\pi^-)/dz$ decreases with increasing $z$, 
we take a conservative estimate by considering 
the measurement in the range of $0.95 < z < 0.98$; 
this gives $\sigma(e^+e^- \to \pi^+\pi^-) \simeq 0.3$ pb, 
thus leading to $\simeq 1.5 \times 10^7$  events for 
an integrated luminosity of 50 ab$^{-1}$ at Belle II.
Since we only consider the barrel region of the ECL, 
we adopt the mis-identification rate 
for $\pi^-$ as $e^-$ 
in the high $p_T$ region, 
which is $\sim 3\times 10^{-5}$ \cite{Milesi:2020esq}. 
To determine the probability 
for a $\pi^+$ to deposit $<5\%$ of its energy in the ECL, 
we carry out GEANT4 simulations 
and find that this probability is $\lesssim 10^{-5}$. 
Taking into account these two factors, backgrounds 
due to the $\pi^+ \pi^-$ final state 
are $\simeq 4.5\times 10^{-3}$. 
Thus, we expect that backgrounds due to a hadronic pair 
are negligible.

We next discuss backgrounds due to the $\mu^+\mu^-$ final state 
generated in $e^+e^-$ collisions at the primary vertex, 
which has a production cross section 
of $\simeq 500$ pb at $\sqrt{s} = 10.58 ~\rm GeV$ in the ECL barrel region, 
thus leading to $\simeq 2.5\times 10^{10}$ events 
for an integrated luminosity of 50 ab$^{-1}$ at Belle II. 
To determine the background, 
we carry out GEANT4 simulations for 
both $\mu^-$ and $\mu^+$ in ECL. 
Because the energy resolution of ECL is $\delta E/E \simeq 2\%$ 
\cite{Belle-II:2018jsg}, 
we determine the probability for a $\mu^-$ to mimic an $e^-$ in ECL 
by computing the probability for a $\mu^-$ to deposit  
$>$ 98\% of its energy in ECL; 
we find   
the probability to be $\lesssim 10^{-7}$ 
by using our GEANT4 simulations. 
We note that one of the processes in which 
a $\mu^-$ can mimic an $e^-$ is one 
where $\mu^-$ emits a very hard and collinear radiation.
The probability for a $\mu^+$ to deposit $<5\%$ of its 
energy in ECL is found to be $\simeq 90\%$. 
To estimate the KLM veto efficiency on the $\mu^+$, 
we adopt the KLM veto efficiency on photon, 
which is $\sim 4.5\times 10^{-4}$ \cite{BaBar:2008aby}. 
The backgrounds are reduced to a single event, 
after taking into account the above three factors. 
Given that KLM is 
a muon detector, we expect its veto efficiency for muons to surpass that for photons. 
Consequently, we anticipate that backgrounds 
due to $\mu^+ \mu^-$ pairs will be negligible.

Backgrounds can also arise from the $e^+ e^- e^+ e^-$ final state, 
where a pair of $e^+$ and $e^-$ is outside the acceptance of detectors. 
Another background can arise from the $\tau^+ \tau^-$ final state, 
where 
$\tau^+$ decays into $e^+$/$\mu^+$/$\pi^+$ 
and 
$\tau^-$ decays into $e^-$. 
However, 
due to energy carried away from the other particles, 
these two types of backgrounds do not exhibit the expected electron and positron momenta characteristic of Bhabha scattering events 
and can be effectively vetoed via track measurements in the CDC.

Another potential background  arises 
when a pair of muons is produced 
in collisions between $e^+$ and 
the ECL, via 
the bremsstrahlung process: $e^++N\to e^+ +N+\mu^++\mu^-$, 
which has a cross section of $\sim$300 nb. 
This leads to $\sim$3400 events 
with $6\times 10^{11}$ 
positrons hitting the ECL target. 
Once again, we adopt  
the KLM veto efficiency on photon 
as the efficiency on muon, 
which is $\sim 4.5\times 10^{-4}$ \cite{BaBar:2008aby}; 
applying this veto efficiency to both muons 
makes this type of background negligible.

\section{Sensitivity on the scalar mediator model}
\label{sec:scalar}

In this section we investigate the Belle II sensitivity 
on an alternative DM model
from the ``disappearing positron track'' channel. 
Thus, we consider the scalar mediator model 
with the following interaction Lagrangian 
\be
 {\cal L}_{\rm int} = \phi 
    (\epsilon  e \bar{\ell} \ell 
    + g_\chi \bar \chi \chi), 
\ee
where $\phi$ is the scalar mediator  
with mass $m_{\phi}$, 
and $\ell$ is the SM charged lepton. 
Similar to the invisible dark photon case, 
we assume $m_\chi=m_{\phi}/3$ and $g_\chi \gg e \epsilon$
such that $\phi$ decays dominantly into DM with the  
decay width 
\be
\Gamma_\phi 
=\frac{m_{\phi}\alpha_D}{2}
\left(1-4\frac{m_\chi^2}{m_{\phi}^2} \right)\sqrt{1-\frac{4m_\chi^2}{m_{\phi}^2}}.
\label{eq:scalardecay}
\ee

\begin{figure}[htbp]
\begin{centering}
\includegraphics[width=0.45 \textwidth]{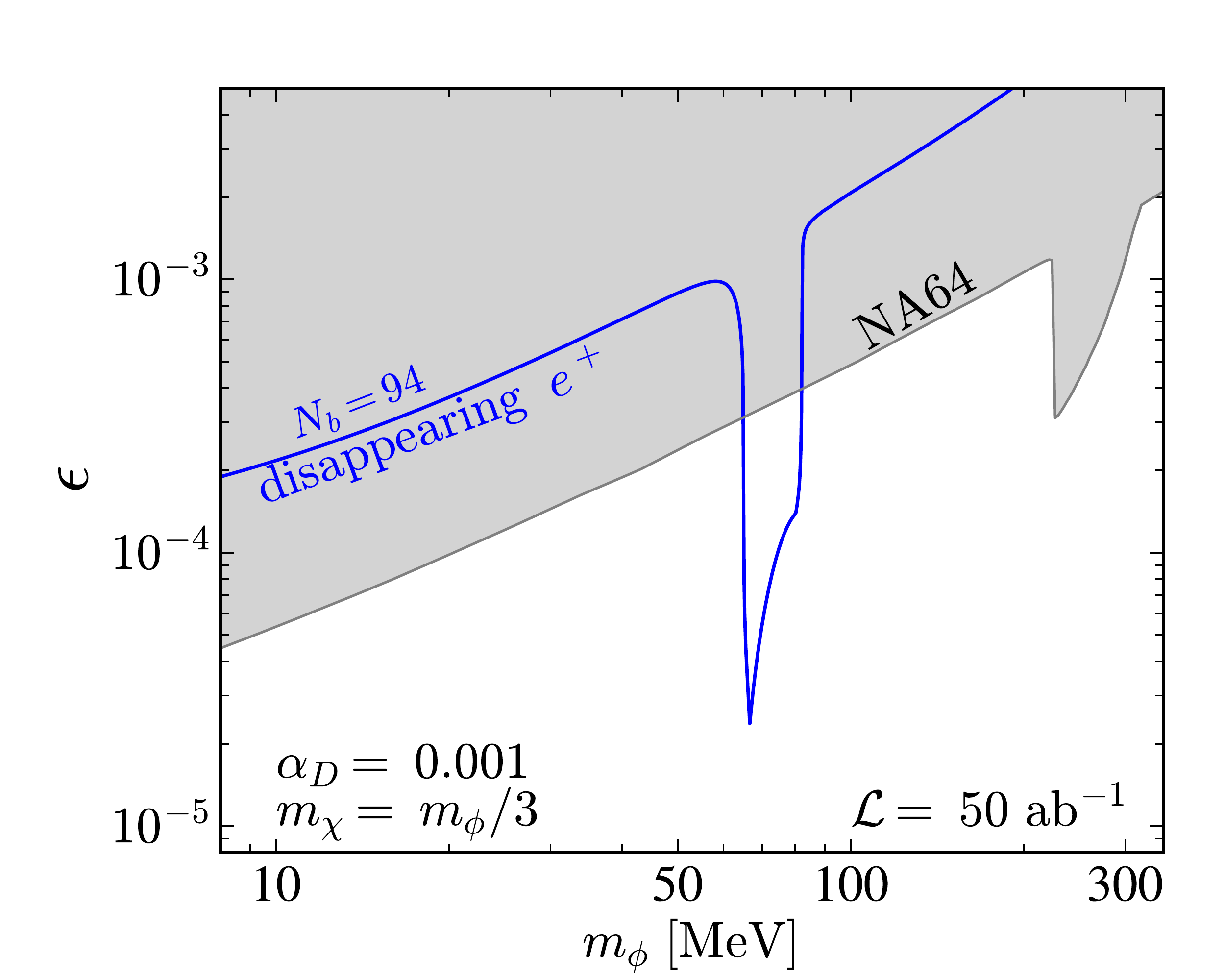}
\caption{Belle II sensitivity on 
the coupling parameter $\epsilon$  
as a function of the scalar mediator mass $m_\phi$
from the ``disappearing positron track'' with $N_b=94$ 
and ${\cal L} = 50 ~{\rm ab}^{-1}$. 
Here we fix $\alpha_D=0.001$ and $m_\chi=m_{\phi}/3$. 
For the NA64 constraint, 
we adopt the limits from Ref.~\cite{NA64:2021xzo} due to the bremsstrahlung process, 
and use the method in appendix \ref{sec:NA64calc} 
for the resonance region.}
\label{fig:BelleIIscalar}
\end{centering}
\end{figure}

We follow the same procedure in section \ref{sec:results} 
to compute the Belle II sensitivity on the scalar mediator model 
from the ``disappearing positron track'' channel, 
which is shown in Fig.~(\ref{fig:BelleIIscalar}). 
Similar to the invisible dark photon model, 
the sensitivity on the scalar mediator model 
is also significantly enhanced in the resonance region near 
$m_{\phi} = 66~\rm MeV$. 
The sensitivity in the resonance region
is dominated by the annihilation process, 
$e^+ e^-_{A} \to \phi \to \chi \bar{\chi}$, 
which 
has the following cross section: 
\be
\sigma_S =\frac{e^2 \epsilon^2 \alpha_D}{4}
\frac{s-4m_\chi^2}{(s-m_{\phi}^2)^2+\Gamma_{\phi}^2 m_{\phi}^2}
\sqrt{1-\frac{4 m_\chi^2}{s}}. 
\label{eq:annxsec-S}
\ee
The best limit on the coupling parameter in the 
resonance region is 
$\epsilon \simeq 2.4 \times 10^{-5}$, 
which is slightly weakened as compared to the dark photon case. 
We note that the change of the $\epsilon$ limit can be 
explained by the fact that 
in the narrow width approximation, 
the new physics signal is proportional to $(2J+1)\Gamma(M\to e^+ e^-)$, 
where $J$ is the spin of the mediator $M$.

\section{Sensitivity of the radiative Bhabha scattering process}
\label{sec:RBS}

\begin{figure}[tbhp]
\begin{centering}
\includegraphics[width=0.5 \textwidth]{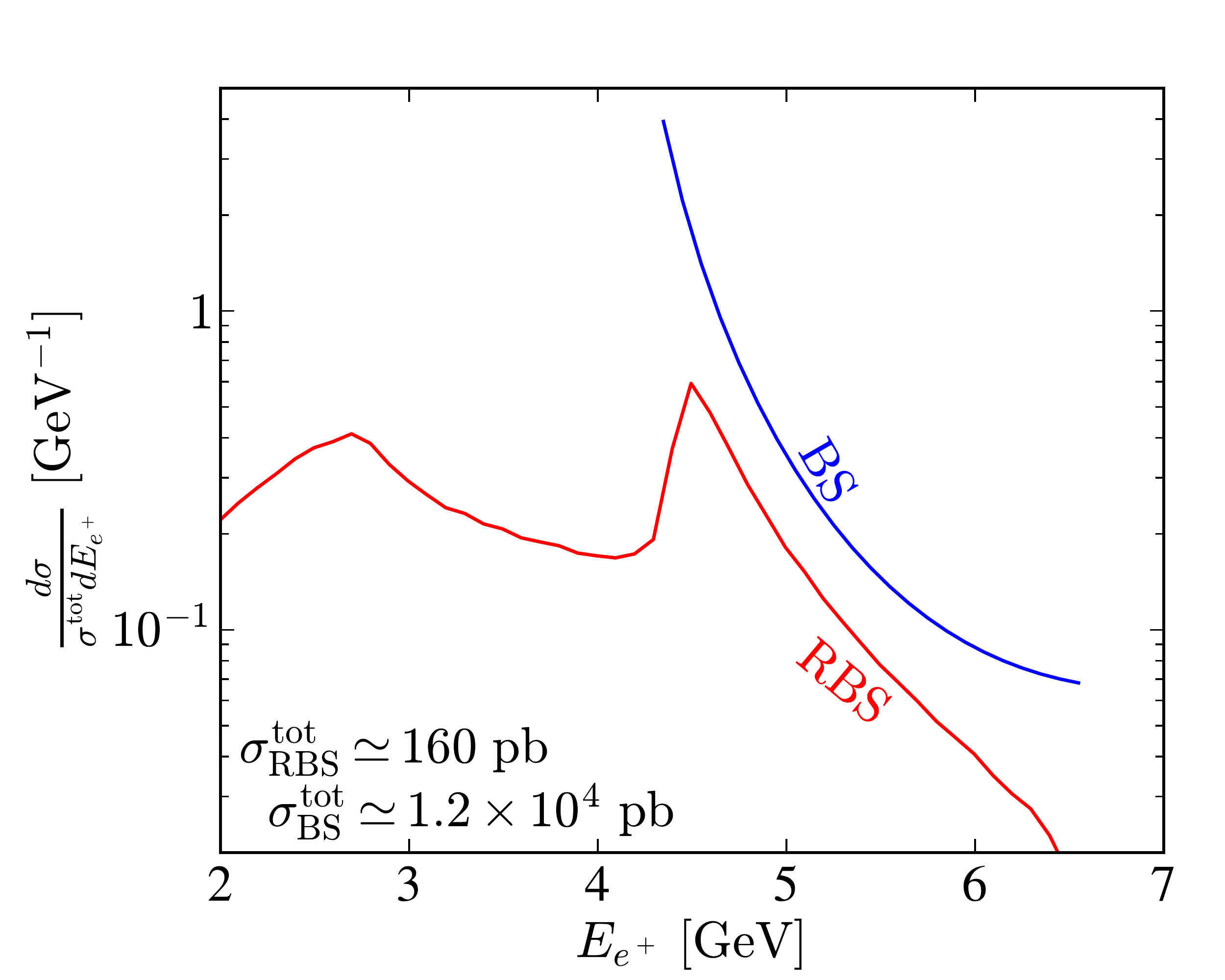}
\caption{The normalized positron energy spectra of both the BS 
($e^+e^-\to e^+e^-$)
and the RBS 
($e^+e^-\to e^+e^-\gamma$) processes, 
where $\sigma^{\rm tot} \simeq 160$ ($1.2\times 10^{4}$) pb 
is the total cross section for the RBS (BS) case. 
The BS case is given by Eq.~\eqref{eq:xsecBha:lab}.
The RBS case is obtained from MadGraph simulations, 
where all the three final state particles are required 
to be within the ECL barrel region 
and have energy above 2 GeV; 
moreover, the photon is required to have a angular distance 
$\Delta R>0.1$ from both the electron and the positron.
} 
\label{fig:RBdiffXsec}
\end{centering}
\end{figure}

In this section we discuss the potential sensitivity of the ``disappearing positron track'' channel 
for positrons generated by the radiative Bhabha scattering (RBS) process, 
$e^+e^-\to e^+e^-\gamma$.
Although the RBS cross section is expected to be smaller by a factor of $\alpha \simeq 1/137$ 
than that of the Bhabha scattering (BS) process, 
positrons from the RBS process exhibit a broader energy range, 
as shown in Fig.~(\ref{fig:RBdiffXsec}), 
which thus have the potential to probe different parameter 
space in the resonance region via the annihilation process.

As discussed in section \ref{sec:channel}, 
the missing energy in the positron track can be 
obtained by comparing the momentum of the positron inferred from 
the CDC track with the energy deposition in the ECL. 
In the BS events, the energy of the positron can be further cross-checked by 
its polar angle and the measurement of the electron. 
In contrast, there is no simple relation between the positron energy 
with its polar angle. 
But one can still 
reconstruct the energies and momenta of all the three final state particles 
and then use the momentum conservation 
to cross-check the measurement on the positron. 
Note that the 
energy resolution of ECL is $\sigma_E/E=1.6\%$ $(4\%)$ at 8 GeV (100 MeV), 
and the angular resolution of ECL at high (low) energies 
is 3 (13) mrad \cite{Belle-II:2018jsg}. 
Thus the angular information of the photon can be well-measured by the ECL.

\begin{figure}[tbhp]
\begin{centering}
\includegraphics[width=0.5 \textwidth]{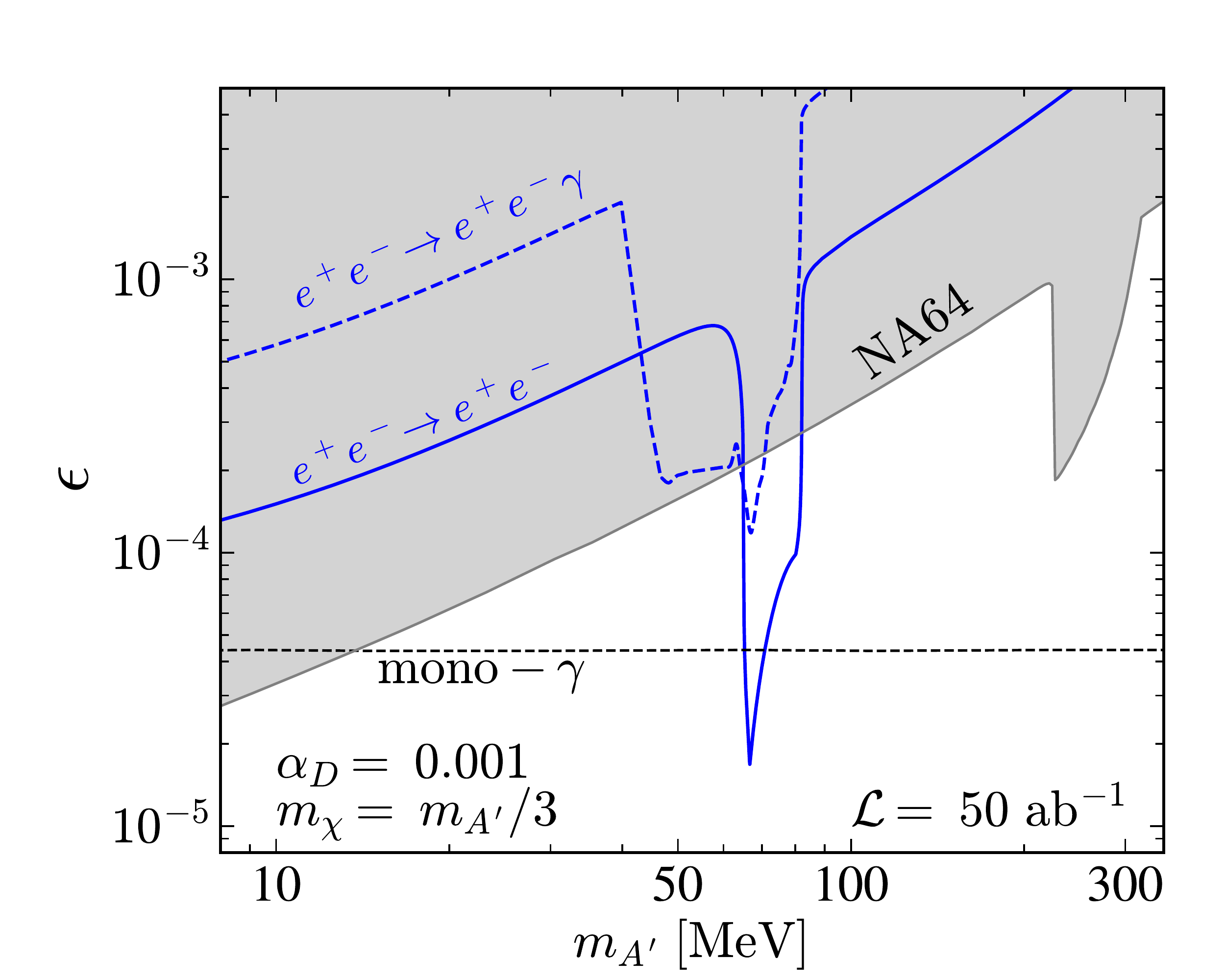}
\caption{Belle II sensitivities on 
the coupling parameter $\epsilon$ 
with $50 ~{\rm ab}^{-1}$ 
integrated luminosity 
as a function of the DP mass $m_{A'}$
from the ``disappearing positron track'' channel  
for the Bhabha scattering process with $N_b=94$ (blue solid) 
and for the radiative Bhabha scattering process with $N_b=1$ (blue dashed). 
Here we fix $\alpha_D=0.001$ and $m_\chi=m_{A'}/3$. 
Also shown are the 
Belle II sensitivity from the mono-photon channel 
with $50 ~{\rm ab}^{-1}$ 
(black dashed), 
and the NA64 constraints (gray shaded region).} 
\label{fig:DPRB}
\end{centering}
\end{figure}

To compute the cross section and the positron energy spectrum in the RBS process, 
we simulated $10^6$ RBS ($e^+e^-\to e^+e^-\gamma$) 
events by using {\sc Madgraph} \cite{Alwall:2014hca}. 
In our simulation, we require each of the three final state particles 
to have $E>2~\rm GeV$ and $32.2^\circ<\theta<128.7^\circ$ in the lab frame; 
we further require that the photon does not appear in the 
vicinity of both the electron and the positron such that 
the angular distance 
$\Delta R_{\gamma e^\pm}$ is greater than 0.1. 
Under these detector cuts, we find that 
the total cross section is $\simeq$160 pb, 
corresponding 
to $\simeq 8\times10^9$ events with a luminosity of 50 ab$^{-1}$. 
To compute the number of the signal events in the invisible dark photon model, 
we use Eq.~\eqref{eq:NDPann} and Eq.~\eqref{eq:NDPbrem}, 
where $d\sigma_{B}/dE$ is obtained by the RBS events 
in our simulation, as shown in Fig.~(\ref{fig:RBdiffXsec}).
In section \ref{sec:BG}, we have found that 
there are $\simeq 94$ backgrounds due to 
punch-through neutrons/photons for 
the $\simeq 6\times 10^{11}$ BS positrons 
in the ECL barrel region. 
Thus,  
one expect $\simeq 1$ background 
for the $\simeq 8\times10^9$ RBS events. 
Fig.~(\ref{fig:DPRB}) shows the 
90\% CL upper bound on $\epsilon$ from the RBS process, 
where the criterion of $N_s=3$ is used. 
The resonance region for the RBS process expands to the range of  
$40~{\rm MeV}\lesssim m_{A'} \lesssim 80~{\rm MeV}$, 
due to the broader range of the positron energy. 
The limit improves within the mass range of 
$40~{\rm MeV} \lesssim m_{A'} \lesssim 60~{\rm MeV}$, 
reaching a level comparable to the NA64 constraints.


\bibliography{ref.bib}{}

\bibliographystyle{utphys28mod}

\end{document}